\documentclass[11pt]{article}

\usepackage{latexsym}
\parskip=\medskipamount

\def\a{\alpha}
\def\b{\beta}
\def\C{\raise2pt\hbox{\rm\large$\chi$}}
\def\c{\chi}
\def\d{\delta}
\def\e{\epsilon}
\def\f{\varphi}
\def\g{\gamma}
\def\h{\eta}

\def\k{\kappa}
\def\l{\lambda}

\def\o{\omega}
\def\p{\pi}
\def\q{\theta}
\def\r{\rho}

\def\F{\Phi}
\def\G{\Gamma}

\def\L{\Lambda}

\def\hd{\widehat{\cd}}

\DeclareFontFamily{OT1}{msb}{}{}
\DeclareFontShape{OT1}{msb}{m}{n}
 {  <5> <6> <7> <8> <9> <10> gen * msbm
      <10.95><12><14.4><17.28><20.74><24.88>msbm10}{}
\DeclareMathAlphabet{\bubble}{OT1}{msb}{m}{n}

\def\bR{{\bubble R}}
\def\bZ{{\bubble Z}}

\def\cd{{\cal D}}

\def\cj{{\cal J}}

\def\cl{{\cal L}}
\def\cm{{\cal M}}

\def\cp{{\cal P}}

\def\cy{{\cal Y}}

\def\tPCO{\tilde{\cp}}

\def\half{{1\over 2}}
\def\pa{\partial}

\def\ra{\rightarrow}
\def\tr{{\rm tr}}
\def\Tr{{\rm Tr}}

\def\ad{\dtt{\a}}
\def\da{\dtt{\a}}
\def\bd{\dtt{\b}}
\def\db{\dtt{\b}}
\def\dg{\dtt{\g}}
\def\dd{\dtt{\d}}
\def\dm{\dtt{\mu}}
\def\dn{\dtt{\nu}}

\def\VEV#1{\left\langle #1\right\rangle}        
\def\leftrightarrowfill{$\mathsurround=0pt \mathord\leftarrow \mkern-6mu
        \cleaders\hbox{$\mkern-2mu \mathord- \mkern-2mu$}\hfill
        \mkern-6mu \mathord\rightarrow$}
\def\dvec#1{\vbox{\ialign{##\crcr
        \leftrightarrowfill\crcr\noalign{\kern-1pt\nointerlineskip}
        $\hfil\displaystyle{#1}\hfil$\crcr}}}           
\def\dtt#1{{\buildrel {\hbox{\LARGE .}} \over {#1}}}     

\def\fr#1#2{{\textstyle{#1\over#2}}}             

\def\beq{\begin{equation}}
\def\eeq{\end{equation}}
\def\beqx{\begin{displaymath}} 
\def\eeqx{\end{displaymath}}
\def\beql{\arraycolsep .1em \begin{eqnarray}}
\def\eeql{\end{eqnarray}}

\def\gl#1{(\ref{#1})}

\def\theequation{\thesection.\arabic{equation}}

\catcode`@=11
\@addtoreset{equation}{section}
\@addtoreset{equation}{subsection}
\def\theequation{\ifnum\value{section}=0 \arabic{equation}\ignorespaces
\else \ifnum\value{section}=-1 A.\arabic{equation}\ignorespaces
\else \ifnum\value{subsection}=0 \thesection.\arabic{equation}\ignorespaces
\else \thesection.\arabic{subsection}.\arabic{equation}\ignorespaces
                           \fi
                      \fi
                 \fi}
\catcode`@=12

\def\sM{${\widehat {\cal M}}^+$} 

\def\0#1{{\stackrel{\circ}{#1}}} 
\def\der#1{{\partial \over \partial #1}}

\def\N#1{$N{=}#1$}
\def\square{\kern1pt\vbox
            {\hrule height 0.6pt\hbox{\vrule width 0.6pt\hskip 3pt
 \vbox{\vskip 6pt}\hskip 3pt\vrule width 0.6pt}\hrule height 0.6pt}\kern1pt}

\def\sd{self-dual}
\def\sdy{self-duality}
\def\sp{super-Poincar\'e algebra}
\def\ym{Yang-Mills}

\def\be{\begin{equation}}
\def\ee{\end{equation}}
\def\la#1{\label{#1}}  
\def\arr{\begin{array}{rll}}
\def\ea{\end{array}}
\def\bea{\begin{eqnarray}}
\def\eea{\end{eqnarray}}

\begin{document}


\begin{titlepage}

\noindent
hep-th/9808053
\hfill ITP--UH--15/98 \\

\vskip 1.0cm

\begin{center}

{\Large\bf String-induced Yang-Mills coupling to self-dual gravity~$^*$}\\

\vskip 1.5cm

{\large Chandrashekar Devchand}

{\it Max-Planck-Institut f\"ur Mathematik in den Naturwissenschaften}\\
{\it Inselstra\ss e 22-26,  04103  Leipzig, Germany}\\
{E-mail: devchand@mis.mpg.de}\\

\vskip 0.7cm

{\large Olaf Lechtenfeld}

{\it Institut f\"ur Theoretische Physik, Universit\"at Hannover}\\
{\it Appelstra\ss{}e 2, 30167 Hannover, Germany}\\
{http://www.itp.uni-hannover.de/\~{}lechtenf/}\\

\vskip 2.5cm

{\bf Abstract}
\end{center}
\begin{quote}
By considering $N{=}2$ string amplitudes we determine the $(2{+}2)$-dimensional
target space action for the physical degrees of freedom:  self-dual gravity and
self-dual Yang-Mills, together with their respective infinite towers of
higher-spin inequivalent picture states.  Novel `stringy' couplings amongst 
these fields are essential ingredients of an action principle for the effective 
target space field theory.  We discuss the covariant description of this theory
in terms of self-dual fields on a hyperspace parametrised by the target space
coordinate and a commuting chiral spinor.  
\end{quote}

\vfill
\hrule width 5.cm
\vskip.1in

{\small \noindent ${}^{*\ {}}$
supported in part by the Deutsche Forschungsgemeinschaft;
grant LE-838/5-2}

\end{titlepage}

\hfuzz=10pt


\section{Introduction}

We have recently presented a covariant description of the physical degrees of
freedom of the \N2 open string in terms of a self-dual Yang-Mills theory on a
{\it hyperspace} parametrised by the coordinates of the $(2{+}2)$-dimensional
target space $x^{\pm\da}$ together with a {\it commuting\/} chiral spinor
$\eta^\pm$ \cite{dl1}.  The infinite tower of massless string degrees of
freedom, corresponding to the inequivalent pictures (spinor ghost vacua) of 
this string~\cite{LS}, are compactly represented by a hyperspace generalisation
of the prepotential originally used by Leznov~\cite{leznov,parkes} 
to encode the dynamical degree of freedom of a \sd\ Yang-Mills (SDYM) theory. 
The generalised hyperspace Leznov Lagrangean yields an action describing the 
tree-level \N2 open string amplitudes~\cite{dl1}.  This description reveals 
the symmetry algebra of the space of physical states to be the {\it Lie-algebra 
extension} of the Poincar\'e algebra \cite{ac} obtained from the \N1 \sp\ 
by changing the statistics of the Grassmann-odd (fermionic) generators. 
Picture-raising~\cite{FMS,BKL} is thus interpreted as an {\it even\/} variant 
of a supersymmetry transformation.

The purpose of this paper is to investigate whether closed strings allow
incorporation into the above picture.  The physical centre-of-mass mode is
well-known to describe self-dual gravity (SDG) in $(2+2)$ dimensions
\cite{OVold}.  The effect of inequivalent picture states has, however, 
hitherto not been taken into account. As for the open case, the closed sector
physical state space consists of an infinite tower of massless picture-states 
of increasing spin~\cite{BL1,BL2}.  In particular the scattering of open 
strings with closed strings determines a particular coupling of the SDG tower 
of picture states with the SDYM tower~\cite{marcus}.  Motivated by our 
previous results for the open string sector~\cite{dl1}, we first (in section 2)
set up the general framework of curved-hyperspace self-duality, in the 
expectation that it underlies the full (open ${+}$ closed) \N2 string dynamics.
This involves a generalisation of the formalism previously developed to study 
self-dual gravity \cite{revisited} and self-dual supergravity \cite{sdsg} 
to the hyperspace introduced in \cite{dl1}.  Our formalism is basically a 
field-theoretical variant of the twistor construction.  The dynamical degrees 
of freedom of hyperspace self-duality are seen to be encoded in a hyperspace 
variant of Plebanski's `heavenly' equation~\cite{plebanski}.  

Gauge covariantising the construction yields a curved hyperspace variant of 
the Leznov equation as well.  These two equations, however, do not provide 
a complete description of the effective \N2 string dynamics, for perusal of 
string scattering amplitudes (section 3) reveals further couplings between the 
gravitational and gauge degrees of freedom.  Taking these into account yields 
an effective target space action (section 4) for the two infinite towers of 
target space fields.  We discuss the hyperspace-covariant description of this 
action and write down homogeneous hyperspace equations of motion.  Finally,
a consistent truncation is performed (section 5) to multiplets of 9 fields from
the Plebanski tower and 5 from the Leznov tower. Their combined action is 
rather reminiscent of the maximally helicity-violating projection of 
(non-self-dual) light-cone $N{=}8$ {\it super\/}gravity plus $N{=}4$ 
{\it super\/} Yang-Mills, with the replacement of the fermionic chiral 
superspace coordinate by a commuting spinor.

\vfil\goodbreak


\section{Self-dual gravity picture album}

Consider a {\it self-dual chiral hyperspace\/} \sM\ with coordinates 
$ \{ x^{\a\dm},  \eta^\a \}$, where
$\eta^\a $ is a {\it commuting\/} spinor
and 
$x^{\a\dm}$ are standard coordinates on $\bR^{2,2}$.  
As for self-dual superspaces, only half the global tangent space group 
$SO(2,2)\simeq SL(2,\bR)\times SL(2,\bR)$ is gauged. One of the 
world indices is therefore identical to the corresponding tangent index
(denoted by early Greek indices $\a,\b,\g,$ etc.)
and only the dotted index has `world' and `tangent' variants. The components
of the spinor $\eta^\a$ therefore do not transform under space diffeomorphisms.
Covariant derivatives in the chiral hyperspace therefore take the form
\be         
\cd_{\a}\ =\  \pa_{\a}\ +\ E^{\b \dm}_{\a }\pa_{\b \dm}\
+\ \o_{\a}\  \qquad,\qquad
\cd_{\a \da}\ =\ E^{\b \dm}_{\a \da}\pa_{\b \dm}\  
+\ \o_{\a\da} 
\quad,\ee
with the partial derivatives $\ \pa_{\a} \equiv \der{\eta^{\a}}\ $ and 
$\pa_{\a \dm} \equiv \der{x^{\a \dm}}\ $.  
The components of the spin connection $( \o_{\a}, \o_{\a\da})$ are determined 
in terms of the vielbein fields in virtue of zero-torsion conditions. We choose 
them in a {\it self-dual gauge}, $\o_{\a}=(\o_{\a} )_\db^\dg \G^\db_\dg$ and 
$\o_{\a\da} = (\o_{\a\da} )_\db^\dg \G^\db_\dg$ , i.e.
taking values in the Lie algebra of the gauged $SL(2,\bR)$.
They therefore act on dotted tangent space indices.
Thus restricting the local part of the tangent space group
to half of the Lorentz group is (gauge) equivalent to imposing self-duality 
on the corresponding curvatures, viz.,
\be\arr
     [\cd_{\a},  \cd_{\b}]\ &=&\ \e_{\a\b} R \\[5pt]
     [\cd_{\a}, \cd_{\b \db}]\ &=&\  \e_{\a\b} R_{ \db}  \\[5pt]
     [\cd_{\a \da}, \cd_{\b \db}]\ &=&\  \e_{\a\b} R_{\da\db} 
     \quad.\la{cvv1}
\ea\ee
With the undotted indices thus `de-gauged', we can proceed in analogy to the 
\ym\ case \cite{dl1} and enlarge \sM\ to a harmonic space with coordinates
$\{ x^{\pm\dm}, \eta^{\pm},  u^\pm_\a \} ,$
where $ x^{\pm\dm} = u^\pm_\a x^{\a\dm},$
and $\eta^{\pm} = u^\pm_\a \eta^{\a} $. 

The equations \gl{cvv1} are  equivalent to the following
curvature constraints:
\bea 
  &      [\cd^+ , \cd^+_\db ]\ =\ 0\quad  ,\qquad 
        [\cd^+_\da , \cd^+_\db ]\ =\ 0   \la{c++}\\[5pt]       
   &  [\cd^- , \cd^-_\db ]\ =\ 0\quad  ,\qquad 
    [\cd^-_\da , \cd^-_\db ]\ =\ 0  \la{c--}\\[5pt]       
& [\cd^+ , \cd^- ]\ =\ R \quad  ,\qquad 
[\cd^+ , \cd^-_\db ]\ =\ [ \cd^+_\db , \cd^- ]\ =\ R_\db \quad  ,\qquad 
 [\cd^+_\da , \cd^-_\db ]\ =\ R_{\da\db}\la{c+-} \quad.
\la{cvv2}\eea
These allow, by the usual Frobenius argument, the choice of an analytic 
{\it Frobenius frame}  in which $\cd^+ , \cd^+_\db$ are flat. In this frame,
diffeomorphism and Lorentz invariances are determined in terms of analytic
(independent of $x^{+\dm}, \eta^+$) degrees of freedom. We do not transform the 
harmonic variables $u^\pm_\a$\,. This facilitates the application to \N2 string theory,
which requires a fixing of the complex structure and the use of 
corresponding light-cone variables. Moreover, let us choose the 
transformation parameters to be independent of the spinorial variables
$\eta^{\pm}$. We thus consider the following action of the local group of
infinitesimal diffeomorphisms
\be 
    \d\, x^{+\dm}\ =\ \l^{+\dm}(x^{+\dm},u^\pm)\quad ,\qquad  
    \d\, x^{-\dm}\ =\ \l^{-\dm}(x^{+\dm},x^{-\dm}, u^\pm)\quad ,\qquad  
    \d\, \eta^\pm\ =\ 0 \quad. 
\ee
The gauge choice $\cd^+=\pa^+$ and $\cd^+_\db=\pa^+_\db$  is tantamount to
the following relationship between the non-analytic parameter  $\l^{-\dm}$
and the analytic parameter of local $sl(2,\bR)$ transformations 
$\l_\da^\db = \l_\da^\db(x^{+\dm},u^\pm)$:
\be  
\pa^+_\da \l^{-\dm}\ =\ -\, \l_\da^\dm \quad.
\la{gauge1}\ee
In this frame the covariant derivatives take the form
\be\arr
\cd^+ &=& \pa^+ \\[5pt]
\cd^+_{\da}&=& \pa^+_{\da} \\[5pt]
\cd^- &=& -\pa^-\ +\ E^\dm\pa^-_\dm\ +\ E^{--\dm}\pa^+_\dm\ 
      +\ \o^- \\[5pt]
\cd^-_{\da}  &=& -E_\da^\dm \pa^-_{\dm}\ +\ E^{--\dm}_\da \pa^+_{\dm}\ 
                   +\ \o^-_\da \quad,
\la{D_an}\ea\ee
where the vielbein fields transform as:
\be\arr
\d\, E^\dm     &=&  E^\dn \pa^-_\dn \l^{+\dm}\\[5pt]
\d\, E^{--\dm} &=& \cd^- \l^{-\dm}\\[5pt]
\d\, E_\da^\dm &=& \cd^-_\da \l^{+\dm} + \l_\da^\db E_\db^\dm \\[5pt]
\d\, E^{--\dm}_\da &=& \cd^-_\da \l^{-\dm} + \l_\da^\db E_\db^{--\dm}\quad.
\la{transf}\ea\ee
We note that the fields $E^\dm$ and  $E_\da^\dm$ have transformations depending 
only on analytic transformation parameters. Moreover, in virtue of \gl{c--} and
\gl{c+-} these fields are analytic, satisfying the set of equations
\be\arr
&&\pa^+ E^\dm\ =\ 0\ =\ \pa^+_\da E^\dm \\[5pt]
&&\pa^+ E_\db^\dm\ =\ 0\ =\ \pa^+_\da E_\db^\dm \\[5pt]
&&\cd^- E_\da^\dm + \cd^-_\da E^\dm\ =\ 0 \\[5pt]
&& \cd^-_{[\db} E_{\dg]}^\dm\ =\ 0\quad.
\la{Ee}\ea\ee
We may therefore choose $x^{+\dm}$ such that $E^\dm = 0$ and  
$E_\da^\dm =\d_\da^\dm$\,. 
In this gauge, the relation \gl{gauge1} is supplemented by
\be  
\pa^-_\da \l^{+\dm}\ =\ \l_\da^\dm\quad.
\la{gauge2}\ee
All diffeomorphisms are thus effected by residual Lorentz transformations,
allowing world indices to be freely replaced by tangent indices with an action
of the Lorentz group. In this gauge, the last two equations in \gl{Ee} 
yield the conditions 
\be
(\o^- )^\dd_{\dg}\ =\ 0\quad,\qquad  (\o^-_{[\db} )^\dd_{\dg]}\ =\ 0\quad.
\la{o1}\ee
The former condition is consistent with the $\eta$-independence of the
Lorentz parameters $\l^\db_\da\,$. Clearly, the curvature constraints 
$R=0=R_\da$ follow.
The non-trivial covariant derivatives therefore take the simpler form
\be\arr
\cd^- &=& -\pa^-\  +\ E^{--\dm}\pa^+_\dm\   \\[5pt]
\cd^-_{\da}  &=& -\pa^-_{\da}\ 
                  +\ E^{--\dm}_\da \pa^+_{\dm}\  
                   +\ \o^-_\da \quad.
\la{D_frob}\ea\ee
For the vielbeins appearing here, the torsion constraints implicit in \gl{c--}
and \gl{c+-} yield the equations,
\bea
&&\pa^+ E^{--\dm}\ =\ 0\ =\ \pa^+_\da E^{--\dm} \\[5pt]
&&\pa^+ E_\db^{--\dm}\ =\ 0 \la{an}\\[5pt]
&& \pa^+_\da E_\db^{--\dg}\ =\ (\o^-_\db )_\da^\dg\quad. \la{o2}
\eea
The latter, together with \gl{o1} and the tracelessness of $sl(2,\bR)$
matrices (viz. $(\o^-_\db)_\da^\da=0$), yield an expression for the 
connection component $\o^-_\da$ and vielbein $E^{--\dm}_\da$\,,
\be
\left(\o^-_\da  \right)^\dg_\db\ =\   \pa^+_\da E_\db^{--\dg}\ =\ 
                           \pa^+_{\da} \pa^{+\dg} \pa^+_{\db} F^{----}\quad,
\ee
where the prepotential $F^{----}$ is $\eta^-$-independent, 
$\pa^+ F^{----} =0$\,, so as to satisfy \gl{an}. 
The $\eta^+$-dependence of $E^{--\dm}_\da$ yields the
remaining vielbein, which satisfies the linear equation
\be 
\cd^-_\da E^{--\db}\ =\ -\, \pa^- E^{--\db}_\da\quad.
\ee
The torsion constraint from \gl{c--}, 
\be
\cd^-_{[\db} E^{--\dm}_{\dg]}\ =\ 0\quad,
\ee 
yields, on using an analytic pre-gauge invariance of $F^{----}$, 
the extended Plebanski equation,
\be
\pa^{-\da} \pa^+_\da  F^{----}\ =\  
\fr12\,\pa^{+\da}\pa^{+\db}F^{----}\ \pa^+_{\da}\pa^+_{\db}F^{----}\quad.
\la{P}\ee
Since $F^{----}$ is $\h^-$-independent and transforms in an $\h$-independent 
fashion, it can be thought of as a Laurent expansion in $\h^+$. This equation
therefore encapsulates an infinite tower of equations. Its Lagrangean is of
the compact Plebanski form, with a potential term of the cubic 
Monge-Amp\`{e}re type: 
\be
\cl^{(-8)}_P\ =\  \fr12\, \pa^{-\da}F^{----}\, \pa^+_\da  F^{----}\ +\  
\fr16\,F^{----}\,\pa^{+\da}\pa^{+\db}F^{----}\ \pa^+_{\da}\pa^+_{\db}F^{----}\ .
\la{P_lagr}\ee

So far we have just considered pure self-dual gravity 
in $(2{+}2)$-dimensional chiral hyperspace.
Let us now add self-dual Yang-Mills degrees of freedom 
by gauge covariantising the curvature constraints~\gl{cvv2}. In the analytic 
gauge ($A^+=0=A^+_\da$), this is achieved by `minimally coupling'
Yang-Mills potentials to the negatively charged covariant derivatives,
with coupling constant $g$,
\be\arr
\cd^-  &\rightarrow& \widehat\cd^-\ =\ \cd^-\ +\ gA^- \\[5pt]
\cd^-_{\da} &\rightarrow& \widehat\cd^-_{\da}\ =\  \cd^-_{\da}\ +\ gA^-_{\da}
\quad.
\la{D_gym}\ea\ee
The components of the gauge potentials take values in the Lie algebra of 
the gauge group and have the gauge transformations 
\be\arr
A^- &\ra& \L\, A^-\, \L^{-1}\ -\  \fr1g\, \cd^- \L \,\L^{-1}\\[5pt]
A^-_{\da} &\ra& \L\, A^-_{\da}\, \L^{-1}\ -\  \fr1g\,\cd^-_{\da}\L\,\L^{-1}
\quad,
\ea\ee
with analytic parameter $\L = \L(x^+,\eta^+,u)$ taking values in the gauge 
group.

The coupled gravity-Yang-Mills \sdy\ conditions thus take the form of the 
curvature constraints
\bea 
&[\hd^-_\da , \hd^-_\db ]\ =\ 0 \la{c1vv}\\[5pt] 
&[\hd^- , \hd^-_\da ]\ =\ 0 \la{c1sv}\\[5pt] 
&[\pa^+ , \hd^- ]\ =\  F \quad,\qquad 
[\pa^+ , \hd^-_\da ]\ =\    [ \pa^+_\da , \hd^- ]\ =\ F_\da \la{c2}\\[5pt] 
&[\hd^+_\da , \hd^-_\db ]\ =\   R_{\da\db}\ +\ F_{\da\db}\la{c3}\quad.
\eea
Here, $F_{\da\db}(x,\eta)$, resp. $R_{\da\db}(x,\eta)$, are symmetric and have 
the corresponding  $\bR^{2,2}$ \ym, resp. Weyl, self-dual curvatures as 
their evaluations at $\eta=0$. The fields 
$F\,,\, F_\da\,$ and $F_{\da\db}\,$ take values in the gauge algebra. 

In virtue of \gl{c2} and \gl{c3} we have the expressions
\bea
&A^-\ =\ \pa^+ \F^{--}\quad,\quad A^-_{\da}\ =\ \pa^+_{\da} \F^{--}\\[5pt]
&F\ =\ \pa^+\pa^+ \F^{--}\quad,\quad  F_\da\ =\ \pa^+\pa^+_{\da} \F^{--}
\quad,\quad  F_{\da\db}\ =\ \pa^+_\db \pa^+_{\da} \F^{--}
\eea
in terms of a generalised Leznov prepotential $\F^{--}$. These expressions 
maintain their flat space forms \cite{dl1} since in the analytic gauge the 
positively-charged derivatives remain `flat' in both gauge and gravitational
senses. 
The Plebanski prepotential $F^{----}$ is a scalar under the gauge group; 
its equation remains unmodified by the \ym\ coupling. 
This is consistent with the stress-free self-dual \ym\ field 
{\it not} providing any source for the gravitational field.
The equation for $\F^{--}$, on the other hand, is a
generally covariant version of the Leznov equation obtained from the
gauge-algebra valued part of  \gl{c1vv},
\be
 \cd^{-\da} \pa^+_\da  \F^{--}\  
  +\  \fr{g}{2}\, \left[ \pa^{+\da} \F^{--}\  ,\ \pa^+_\da \F^{--} \right]\
 =\ 0 \quad.
\ee
More explicitly,
\be
 \pa^{-\da} \pa^+_\da  \F^{--}\  
  =\ \pa^{+\da} \pa^{+\db} F^{----}\  \pa^+_\da \pa^+_\db \F^{--}\ 
  +\ \fr{g}{2}\, \left[ \pa^{+\da} \F^{--}\  ,\ \pa^+_\da \F^{--} \right]\quad.
\la{L}\ee
This is the equation which determines the effective dynamics on $\bR^{2,2}$ of
the residual vector potential
$A^-_\da = \pa^+_\da \F^{--}$. The remaining equation for $\F^{--}$, the one
arising from the gauge-algebra part of \gl{c1sv}, determines the 
$\eta^+$-evolution of $A^-_\da$ from its $\eta^+=0$ `initial data', namely,
\be
\pa^- A^-_\da\ =\  \pa^-_\da \pa^+ \F^{--}\ +\  
                   E^{--\db}\ \pa^+_\da \pa^+_\db  \F^{--} \ -\  
                   E_\da^{--\db}\ \pa^+ \pa^+_\db  \F^{--}\ +\  
                  g\, \left[ \pa^+ \F^{--}\  ,\ \pa^+_\da \F^{--} \right]\ .
\ee
It may be noted that the combined system of equations \gl{P} and \gl{L}
cannot be derived from an action principle, since their mutual coupling
appears only in \gl{L}. In the next section, we shall see that new
string-induced couplings provide a remedy.

The hyperspace fields $F^{----}(x,\eta)$ and $\F^{--}(x,\eta)$ can clearly be 
thought of as $\bR^{2,2}$ fields, taking values in the infinite-dimensional 
algebra spanned by polynomials of $\eta^\a$. Such algebras have been 
investigated in~\cite{v}. In particular, consistent 
higher-spin free-field equations were shown to arise as components of 
zero-curvature conditions for connections taking values in such algebras. 
It remains to be seen whether our equations allow interpretation as 
interacting variants of these free-field equations.


\section{Open and closed string amplitudes}

Having obtained the dynamical equations \gl{P} and \gl{L} for the self-dual
hyperspace gravitational and gauge degrees of freedom, we are ready to ask
whether these provide a correct description of $N{=}2$ string dynamics.  By
considering scattering amplitudes, we shall see that these naive equations
require modification which, moreover, yields equations derivable from an 
action principle. The required modification includes `stringy' contributions 
which vanish in the infinite string tension limit.

Since we would like to describe in particular the coupling of self-dual
Yang-Mills to self-dual gravity (in $2{+}2$ dimensions),
let us consider a general (mixed) scattering amplitude involving
$n_c$ closed and $n_o$ open $N{=}2$ strings.
Each such amplitude has a topological expansion in powers of 
the open string coupling~$g$ and in powers of 
a phase given by the angle~$\q$ of the spectral flow. 
The expansion is governed by the world-sheet instanton number~$c\in\bZ$
and the world-sheet Euler number $\c=2-2h-b-x$ in the presence of
$h$ handles, $b$ boundaries and $x$ cross-caps.
Introducing the `spin'
\beq
J\ =\ 2n_c + n_o -2\c \ =\ 2n_c + n_o - 4 + 4h + 2b + 2x
\la{spin}
\eeq
one finds \cite{berk1,LS,dl1} for any given choice of $(n_c,n_o)$, 
the amplitude,
\beq
A\ =\ \sum_J\ g^J \ A_J \
   =\ \sum_{J,c} \left( 2J \atop J{+}c \right)\ g^J \ 
      \sin^{J-c}\fr{\q}{2}\ \cos^{J+c}\fr{\q}{2}\ A_{J,c}
\la{topexp}
\eeq
where $A_{J,c}$ is a correlator of vertex operators~$V$ 
on a world-sheet of fixed topology, integrated over all moduli.
Clearly, $J$ runs upwards in steps of two, 
starting from $J_{\rm min}=2n_c+n_o-2-2\d_{n_o,0}$.
Due to unbalanced spinor ghost zero modes, the instanton sum is
constrained to $|c|\leq J$.
In four dimensions, the open-string coupling~$g$ is just the
(dimensionless) Yang-Mills gauge coupling, while the closed-string
coupling~$g^2$ is related to the (dimensionful) gravitational coupling~$\k$
via 
\beq
\k\ \sim\ \sqrt{\a'}g^2
\eeq
where $\a'$ denotes the inverse string tension.

The string coupling~$g$ and the $\q$ angle
change under global $SO(2,2)$ tangent space transformations 
of the target space~\cite{parkes,BL1,dl1}.
We may therefore set $g=1$ and $\q=0$ for convenience.\footnote{
The dependence on $g$ and $\q$ may easily be restored 
by performing an appropriate $SO(2,2)$ transformation.}
As a consequence, only the top instanton number, $c=J$, contributes, i.e.
\beq
A\ =\ \sum_J A_{J,J}
\la{amptop}
\eeq
with $A_{J,J}$ carrying helicity~$J$.\footnote{
We split $so(2,2)=sl(2)\oplus sl(2)'$. 
`Helicity' is the eigenvalue of the non-compact generator of $sl(2)$ 
which we choose to diagonalise.}
Each partial amplitude $A_{J,J}$ contains an integral 
over four moduli spaces~$\cm_i\,$, corresponding to the 
world-sheet supergravitational non-gauge degrees of freedom of 
the metric, the two gravitini, and the Maxwell field.
The respective real dimensions are
\bea
{\rm dim}_\bR \cm_{\rm metric}\ &=&\ 
2n_c+n_o-3\c \ =\ J-\c \nonumber\\
{\rm dim}_\bR \cm^\pm_{\rm gravitino} \ &=&\ 
2n_c+n_o-2\c\pm c \ =\ J\pm c \ 
\buildrel{c=J}\over{\longrightarrow}\ \cases{2J & ($+$)\cr 0 & ($-$)\cr} \\
{\rm dim}_\bR \cm^\pm_{\rm Maxwell}\ &=&\
2n_c+n_o- \c \ =\ J+\c \nonumber
\eea
The gravitini modular integral can be performed and yields $2J$ 
picture-raising insertions $\tPCO^+$ in the path integral~\cite{BKL,BL2}.
For $\c>0$, the Maxwell modular integral is trivial due
to spectral flow invariance.
Likewise, the metric moduli for positive $\c$ reduce to 
the world-sheet positions of the vertex operators.
The final matter-plus-ghost path integral is a superconformal
correlation function of antighost zero modes, picture-raisers, and (canonical) 
local vertex operators which create the string states from the vacuum.

Which string states are to be scattered?
It is known from the relative BRST cohomology of the {\it open} $N{=}2$ string
that its spectrum consists of a single massless physical state 
at each value of the picture charge $(\p_+,\p_-)$ 
labelling inequivalent spinor ghost vacua \cite{JL}.
An internal symmetry group~$G$ is incorporated by requiring these states 
to carry Chan-Paton adjoint representation indices of~$G$. The difference 
$\p_+{-}\p_-$ changes continuously under the action of spectral flow. 
Since the Maxwell modular integration entails an averaging over the parameter
$\r$ of spectral flow, one must identify the equivalent sectors
$(\p_+,\p_-)\sim (\p_+{+}\r,\p_-{-}\r)$.
We shall use the $\p_+{=}\p_-$ representative.
The total picture number $\p\equiv\p_+{+}\p_-$ takes integral values,
and the helicity of the state is $j=1{+}\p/2\in\half\bZ$.
For states of non-zero momentum~$k$, picture-changing can be used
to implement an equivalence relation among all pictures~\cite{JL}.
In this case, a single open string state interacts with itself,
unless the Chan-Paton group~$G$ is abelian.
Such an identification, however, ruins target space covariance,
because picture-raising by $\tPCO^+$ increases not only $\p$ by one,
but also the helicity~$j$ by half a unit.
It is therefore advantageous to distinguish the unique physical states
in different pictures~$\p$.
The canonical ($j{=}0$) open string state resides at $\p{=}-2$.
The corresponding (canonical) vertex operators $V^o_{\p_i=-2}(k_i)$ are
located on the world-sheet boundaries and feed in target space momenta $k_i\,$.

The scattering amplitude does not depend on the positions 
of the $2J$ picture-raisers~$\tPCO^+$; 
this is the statement of picture equivalence~\cite{FMS}.
Hence, we are free to arbitrarily fuse the picture-raisers 
with some of the vertex operators which raises their picture assignments,
$V^o_{-2} \to V^o_{\p>-2}$. 
In this way we arrive at a correlation function of the form~\footnote{
We suppress the additional appearance of $J{-}\c$ conformal antighost and
$J{+}\c$ Maxwell antighost insertions, which balance the ghost charges of the 
{\it local\/} vertex operators and are used to transform them to 
{\it integrated\/} ones.}
\beq
\VEV{\ V^o_{\p_1}(k_1)\ V^o_{\p_2}(k_2)\ \ldots\ V^o_{\p_{n_o}}(k_{n_o})\ }
\la{opencorr}\eeq
with the picture or helicity selection rule
\be\arr
\p_{\rm tot}\ &=&\ \sum_{i=1}^{n_o} \p_i\ =\ -2n_o + 2J\ =\ -4\c \\[5pt]
j_{\rm tot}\ &=&\ \sum_{i=1}^{n_o} j_i\ =\ \half 2J\ =\ J \quad.
\la{select1}\ea\ee
The first line follows from the second since $\p=-2{+}2j$.
Since the correlator \gl{opencorr} is invariant under picture-changes, 
its value cannot depend on the distribution of $\{\p_i\}$ (or $\{j_i\}$), 
provided the selection rules~\gl{select1} hold. 
A preferred arrangement of picture charges is
\beq
\VEV{\ V^o_{-4\c}(k_1)\ V^o_{0}(k_2)\ \ldots\ V^o_{0}(k_{n_o})\ }
\la{pref}\eeq
corresponding to $j_i=1-2\c\d_{i1}$.

In order to repeat this analysis for the closed string, we note that
the semi-relative BRST cohomology of the closed $N{=}2$ string also consists
of a single massless (now, color-singlet) physical state 
at any value of the picture charge $(\p_+,\p_-)$~\cite{JL}. In principle,
one could consider independent left- and right-moving picture charges; 
however, they need to be identified in the semi-relative construction.
Moreover, since the coupling to open strings via world-sheet boundaries 
enforces a left--right relation for global properties, 
there exists only a single set of picture charges.
Compared to the open string case, the only alteration then is a different
canonical picture for the physical excitation, namely $\p{=}-4$ with $j{=}0$. 
This modifies the picture--helicity relation to $\p=-4{+}2j$ but does not
change the selection rules~\gl{select1}.
Of course, closed-string vertex operators $V^c_\p$ are to be inserted
in the interior of the world-sheet.
A convenient distribution of picture charges among the vertex operators
inside a closed-string correlator is obtained from \gl{pref}
by replacing the labels `$o$' by `$c$'.

Any theory of open strings generates intermediate closed strings at the
loop level. The most general setup contains them already at tree level,
i.e. as external closed-string states. Thus, it is of interest to consider
{\it mixed amplitudes}, which describe scattering processes involving 
open as well as closed string external states.
Having already allowed for handles, boundaries and cross caps of the world
sheet, we simply consider both kinds of vertex operators simultaneously.  
The selection rule
\beq
\p_{\rm tot}\ =\ -4\c 
\qquad\qquad{\rm resp.}\qquad\qquad
 j_{\rm tot}\ =\ J
\la{select2}\eeq
for a mixed correlator
\beq
A_{J,J}^{o\ldots o\ c\ldots c}\ \sim\
\VEV{\ \prod_{i=1}^{n_o} V^o_{\p_i}(k_i)\ 
       \prod_{j=n_o+1}^{n_o+n_c} V^c_{\p_j}(k_j)\ }
\eeq
remains in effect;
and we may again choose all but one external state in the $\p{=}0$ picture.
Using \gl{select2}, we see that, for a given set 
$\{j_1,\ldots,j_{n_o};j_{n_o+1},\ldots,j_{n_o+n_c}\}$
of external state helicities,
the amplitude \gl{amptop} only receives contributions from topologies
with fixed Euler number 
\beq
\c\ =\ (2n_c + n_o - J)/2
\eeq
where $J=\sum_i j_i$.
Because $j_i\in\half\bZ$, there are infinitely many picture-equivalent
and therefore identical amplitudes even at tree level ($\c{>}0$).
Since not much is known yet about loop amplitudes of $N{=}2$ strings,
let us collect the results on the $\c{=}2$ and $\c{=}1$ amplitudes.

\noindent\underline{$\c=2$}\\
The only topology is the sphere, and it does not admit open string legs.
The external helicities sum to $J=2n_c{-}4$. 
It has been shown~\cite{hippmann} that all these amplitudes vanish,
except for the three-point function ($J{=}2$)~\footnote{
Refs.~\cite{OVold,marcus} computed the $A_{2,0}^{ccc}$ component;
the full result was obtained in~\cite{berk1,buckow}.}
\beq
A^{n_c=3,n_o=0}\ =\ A_{2,2}^{ccc}\ =\ 
\sqrt{\a'} \left( k^{++}_{12} \right)^2 \quad.
\la{3cl2}\eeq
Here,
\beq
k_{ij}^{++}\ =\ k_i^{+\ad}\ k_j^{+\bd}\ \e_{\ad\bd}
	   \ =\ \k_i^+ \k_j^+\ \c_{ij} \ =\ -k_{ji}^{++}
\eeq
for lightlike target space momenta 
\beq
k_i^{\a\ad}\ =\ \k_i^\a\ \c_i^\ad \ \in\ \bR^{2,2} \qquad
{\rm and}\qquad
\c_{ij}\ =\ \c_i^\ad\ \c_j^\bd\ \e_{\ad\bd}\ =\ -\c_{ji} \quad.
\eeq
Note that 
\beq
k_{12}^{++}\ =\ k_{23}^{++}\ =\ k_{31}^{++}
\eeq
for massless three-point kinematics since $k_1{+}k_2{+}k_3=0$.
Thus, \gl{3cl2} is totally symmetric in the external legs,
although every helicity assignment (for example, $(-2,2,2)$)
is necessarily asymmetric.
The inverse string tension, $\a'$, must enter \gl{3cl2} on dimensional grounds.

\noindent\underline{$\c=1$: three-point}\\
This situation admits a single boundary or a cross-cap, and is therefore
still interpreted as tree level. 
The cross cap leads to the real projective plane, which only appears
for (unoriented) closed string scattering, i.e. in $A^{ccc}$.
The boundary case is the familiar disk or, equivalently, upper half plane, 
which contributes to all three-string amplitudes.
The results are (see also~\cite{marcus} for the $c{=}0$ parts)
\bea
A_{1,1}^{ooo}\ &=&\ f^{a_1a_2a_3}\ k_{12}^{++} \\
A_{2,2}^{ooc}\ &=&\ \d^{a_1a_2}\ \sqrt{\a'} \left( k_{12}^{++} \right)^2 \\
A_{3,3}^{occ}\ &=&\ 0 \\
A_{4,4}^{ccc}\ &=&\ \g\ \sqrt{\a'}^3 \left( k_{12}^{++} \right)^4 
\la{3point}\eea
where $a_i$ is the adjoint representation Chan-Paton label of the $i$th string 
leg, $f^{a_1a_2a_3}$ are structure constants of the Lie algebra of~$G$,
and $\g$ is a {\it finite\/} numerical constant depending on~$G$.\footnote{
The disc contribution to $A^{ccc}$ involves a Chan-Paton factor 
of~$\tr{\bf1}$ coming from the boundary, 
which the real projective plane does not have.}

It is important to note that for the $N{=}2$ string, 
in contrast to bosonic and ordinary ($N{=}1$) superstrings,
the `higher-order tree' corrections to closed-string scattering are finite.
In the limit of the boundary shrinking to a point, the integrand
of $A^{ccc}$ should yield a (diverging) dilaton propagator at zero momentum 
multiplying $A^{cccc}(k_4{=}0)$ on the sphere.
Obviously, the finiteness of $A^{ccc}$ on the disk
is consistent with the vanishing of the four-point function!
Hence, we do {\it not\/} seem to be forced to take $G=SO(2^{d/2})$ 
in order to cancel infrared divergences.
Nevertheless, it would be interesting to know whether $\g$ can be
made to vanish for some distinguished choice of Chan-Paton group.

As expected, $A_{2,2}^{ccc}\sim(A_{1,1}^{ooo})^2$.
It is instructive to apply an $SO(2,2)$ transformation and restore
the generic $\q$ dependence; for instance
\beq
A_1^{ooo}\ \sim\ \cos^2\fr{\q}{2}\ k_{12}^{++} \
	     +\ 2\cos\fr{\q}{2}\sin\fr{\q}{2}\ i k_{12}^{+-} \
	     +\ \sin^2\fr{\q}{2}\ k_{12}^{--} \quad,
\eeq
with the obvious definition for $k_{12}^{\a\b}$.
This shows again that the interacting $N{=}2$ string 
lives in a broken phase of target space $SO(2,2)$ symmetry. 
The Goldstone modes of the $SO(2,2)\to U(1,1)$ breaking are
precisely the spacetime dilaton and axion fields~\cite{BL1}.
Due to the identity
\beq
k_{ij}^{++}\ k_{ij}^{--}\ =\ k_{ij}^{+-}\ k_{ij}^{-+}
\eeq
for lightlike momenta, the $\c{=}2$ three-point amplitude indeed factorises:
\be\arr
A_2^{ccc}\ &\sim&\ 
		  \cos^4\fr{\q}{2}\ k_{12}^{++}k_{12}^{++} \
		+\ 4\cos^3\fr{\q}{2}\sin\fr{\q}{2}\ i k_{12}^{+-}k_{12}^{++} \
		+\ 6\cos^2\fr{\q}{2}\sin^2\fr{\q}{2}\ k_{12}^{--}k_{12}^{++}
		\\[4pt] && \!\!\!\!\!
		+\ \sin^4\fr{\q}{2}\ k_{12}^{--}k_{12}^{--} \
	        +\ 4\cos\fr{\q}{2}\sin^3\fr{\q}{2}\ i k_{12}^{--}k_{12}^{+-}
		\\[5pt] &\sim&\ \left( A_1^{ooo} \right)^2 \quad.   
\ea\ee
Apparently, 
the question~\cite{BL2} of whether one has a single (joint left-right)
instanton number, or two independent ones (left and right), is irrelevant.

\noindent\underline{$\c=1$: beyond three-point}\\
Tree-level four-point functions are the first place 
to see `stringy' dynamics, 
but they are not easy to compute for mixed 
(i.e. open plus closed string) cases.
Calculations (see \cite{marcus} for the $c{=}0$ piece) have revealed that
\be\arr
A_{2,2}^{oooo}\ &=&\ 0 \\
A_{3,3}^{oooc}\ &=&\ 0 \\
A_{4,4}^{oocc}\ &=&\ 0 ? \\
A_{5,5}^{occc}\ &=&\ 0 \\
A_{6,6}^{cccc}\ &=&\ 0 ?
\ea\la{4point}\ee
where the question marks denote conjectured vanishing amplitudes. 
As argued below, these follow from the assumption that
the target space field theory requires no fundamental quartic vertex,
an expectation from self-dual Yang-Mills plus gravity.
Beyond four-point functions, we only know that the pure open-string
disk amplitude, $A_{n_o-2,n_o-2}^{oo\ldots o}$, vanishes~\cite{hippmann}.


\section{String target space actions}

Knowledge of the $\c{>}0$ three-point string amplitudes allows us
to read off the cubic couplings of the target space action for the
massless open and closed $N{=}2$ string excitations.
We associate string states (resp. their vertex operators) with
space-time fields or their Fourier representatives,
\be\arr
V^o_\pi(k) \ &\Leftrightarrow&\ 
\widetilde\f_{(j)}(k)\ =\ 
\widetilde\f^{\overbrace{--\cdots-}^{2j\rm\;times}}(k) \\
V^c_\pi(k) \ &\Leftrightarrow&\
\widetilde f_{(j)}(k)\ =\ 
\widetilde f^{\overbrace{--\cdots-}^{2j\rm\;times}}(k)\quad, 
\ea\ee
remembering that $j=1{+}\p/2$ for open (and $j=2{+}\p/2$  for closed) string 
states. Fourier transforming to coordinate space (and dropping the tildes)
we find that the $\c{>}0$ three-point functions \gl{3cl2} and \gl{3point}
are reproduced by the target-space Lagrangean density
\bea
\cl_\infty\ &=&  -\ \fr12\sum_{j\in\bZ/2} f_{(-j)}\square f_{(+j)}\ 
+\ \fr{\sqrt{\a'}}{6} \sum_{J=2} 
f_{(j_1)}\ \pa^{+\ad} \pa^{+\bd} f_{(j_2)}\ \pa^+_\ad \pa^+_\bd f_{(j_3)}
\nonumber \\[5pt] 
&& +\ \fr{\g\sqrt{\a'}^3}{6} \sum_{J=4}\ f_{(j_1)}\
\pa^{+\da}\pa^{+\db}\pa^{+\dg}\pa^{+\dd} f_{(j_2)}\
\pa^+_\da \pa^+_\db \pa^+_\dg \pa^+_\dd  f_{(j_3)}
\nonumber\\[5pt]
&& +\ \Tr\biggl\{ -\fr12\sum_{j\in\bZ/2} \f_{(-j)}\square\f_{(+j)}\
+\ \fr16\sum_{J=1}
\f_{(j_1)}\ \Bigl[\pa^{+\ad}\f_{(j_2)}\ ,\ \pa^+_\ad\f_{(j_3)}\Bigr]\biggr.
\nonumber\\[5pt]
&&\qquad\qquad\qquad\qquad\qquad\quad\;\ -\ \biggl.\fr{\sqrt{\a'}}{2}\sum_{J=2}
\pa^{+\ad} \pa^{+\bd} f_{(j_1)}\ \pa^+_\ad \f_{(j_2)}\ \pa^+_\bd \f_{(j_3)}
\biggr\} 
\la{Linfty}\eea
where $\  J\equiv j_1{+}j_2{+}j_3\  $ in the sums, 
and $\ \square = \pa^{-\da} \pa^+_\da\,$.
A field of helicity~$j$ carries a mass dimension equal to $1{-}j$,
so that $\cl_\infty$ has dimension four as required.
The fundamental interactions among $\{\f_{(j)},f_{(j)}\}$ are purely cubic
and of three types, which may be called {\sl three-graviton\/},
{\sl three-gluon\/}, and {\sl gluon-graviton\/} couplings, respectively.
Furthermore, the couplings are independent of the external helicities
as long as these sum to~$J$.

The conjectured vanishing of the higher $n$-point tree-level string amplitudes
(see \gl{4point}) must be reflected in the target space field theory.
In other words, if $\ {\int}\cl_\infty\ $ is the complete space-time action,
it must imply the on-shell vanishing of all tree-level amplitudes beyond the
three-point functions. A non-trivial check, for example, is that iterating 
the fundamental cubic vertices of \gl{Linfty} yields zero for the 
on-shell four-point functions. This was in fact verified
for the pure gluon and the pure graviton cases in~\cite{OVold,buckow}.
Moreover, it is straightforward to extend these results to the mixed four-point
functions as well, with the help of the kinematic relations,
\be\arr && 
{\displaystyle {k_{12}^{++}\ k_{34}^{++} \over s_{12}}\ =\ 
{k_{23}^{++}\ k_{14}^{++} \over s_{23}}\ =\
{k_{31}^{++}\ k_{24}^{++} \over s_{31}} } \\[10pt] &&
k_{12}^{++}\ k_{34}^{++}\ +\ 
k_{23}^{++}\ k_{14}^{++}\ +\ 
k_{31}^{++}\ k_{24}^{++}\ =\ 0 \quad,
\ea\ee
where $s_{ij}=k_{ij}^{[+-]}\ ,\quad s_{ii}=0$ and $\sum_{i=1}^4 k_i=0$.
An inductive argument shows~\cite{MS} that as a consequence all higher 
tree-level $n$-point functions also vanish.
Thus, the absence of higher than cubic vertices 
in $\cl_\infty$ corresponds perfectly with the tree-level string amplitudes 
computed so far.

Of course, our Lagrangean density $\cl_\infty$ contains infinitely many terms. 
It affords, nevertheless, a compact representation in terms of the 
hyperspace functional
\bea
\cl  &=& \fr1{\a'}\ \cl^{(-8)}\ +\  \Tr\ \cl^{(-4)} 
\nonumber\\[8pt] &=& \fr{1}{\a'}\ \Bigl(
  -\ \fr12 F^{----} \square F^{----}\
  +\ \fr16 F^{----}\ \pa^{+\da} \pa^{+\db} F^{----}\ 
                   \pa^+_\da \pa^+_\db F^{----} 
\nonumber\\[4pt] 
&&\quad +\ 
{\textstyle{\g\a' \over 6}}\ (\eta^+ )^{-4}\  F^{----}\
\pa^{+\da}\pa^{+\db}\pa^{+\dg}\pa^{+\dd} F^{----}\
\pa^+_\da \pa^+_\db \pa^+_\dg \pa^+_\dd  F^{----} \Bigr)  
\nonumber\\[4pt]
&&\quad  +\  \Tr\ \Bigl( - \fr12 \F^{--} \square \F^{--}\ 
 +\  \fr16  \F^{--}\ [\pa^{+\da} \F^{--}\ ,\ \pa^+_\da \F^{--}]
\nonumber\\[4pt] &&  \qquad\qquad
 -\ \fr12 \pa^{+\da}\pa^{+\db}F^{----}\ 
                   \pa^+_\da \F^{--}\ \pa^+_\db \F^{--} \Bigr) \quad. 
\la{hyperaction}\eea
The $\fr1{\a'}$ coefficient of $\cl^{(-8)}$ represents the dimensional 
difference between the gravitational and the gauge couplings.

We now claim that the Lagrangean density $\cl_\infty$ is precisely the 
zero-charge homogeneous projection, $ \cl|_0$,  of this inhomogeneous 
combination of the modified Plebanski functional $\cl^{(-8)}$ and the 
modified Leznov functional $\cl^{(-4)}$.
Specifically, since the field $F^{----}$ is independent of $\eta^-$, we may 
think of it as a Laurent expansion in $\eta^+$,
\be
\fr1{\sqrt{\a'}}F^{----}\ =\ 
           \dots + \eta^+ f^{-----} + f^{----} + (\eta^+)^{-1} f^{---} +
           (\eta^+)^{-2} f^{--} + \dots + (\eta^+)^{-8} f^{++++} + \dots
\la{atlasP}\ee
On the other hand, for $\F^{--}$, the essential data is that at $\eta^+=0$. 
We therefore
consider an $\eta^-$ expansion of $\F^{--}$ evaluated at $\eta^+=0$,
\be  \F^{--}\ =\ \dots + (\eta^-)^{-1} \f^{---} + \f^{--} + \eta^- \f^{-} +
               (\eta^-)^2 \f + (\eta^-)^3 \f^{+} + (\eta^-)^4 \f^{++} +
               (\eta^-)^5 \f^{+++} + \dots
\la{atlasL}\ee
The projection $ |_0$ then, is to the respective homogeneous (i.e. zero-charge)
terms in \gl{hyperaction}: 
coefficients of $(\eta^+ )^{-8}$ for $\cl^{(-8)}$ 
and coefficients of $(\eta^- )^p/(\eta^+ )^q\ $, for all $p,q\ge 0$ such that
$p+q=4$, for the remaining terms. 
This charge-homogenising projection yields the homogeneous component 
Lagrangean \gl{Linfty} from the inhomogeneous hyperspace functional $\cl$.

Although the hyperspace functional $\cl$ is not $U(1)$-charge homogeneous,
the question of whether a covariant hyperspace action exists remains open.
Nevertheless, having the above projection in mind, we may indeed write down
homogeneous equations of motion.  Varying $\F^{--}$ yields the generalised
Leznov equation \gl{L} unmodified, whereas varying $F^{----}$ yields 
a modification of the hyperspace Plebanski equation \gl{P},
\bea
\pa^{-\da} \pa^+_\da  F^{----} &=&  
\fr12\ \pa^{+\da}\pa^{+\db}F^{----}\ \pa^+_{\da}\pa^+_{\db}F^{----}
\nonumber\\[4pt]
&& +\ \fr{\g\a'}{2}\  (\eta^+ )^{-4}\Bigl( 
         \pa^{+\da}\pa^{+\db}\pa^{+\dg}\pa^{+\dd}F^{----}\ 
         \pa^+_{\da}\pa^+_{\db}\pa^+_{\dg}\pa^+_{\dd}F^{----}\Bigr)
\nonumber\\[4pt]
&& -\ \fr{\a'}{2}\ (\eta^-)^4\  
         \Tr\ \pa^{+\da}\pa^{+\db}\F^{--}\ \pa^+_{\da}\pa^+_{\db}\F^{--}\quad.  
\la{mP}\eea
Inserting the expansions \gl{atlasP} and \gl{atlasL} yields the infinite set
of Euler-Lagrange equations for $\cl_{\infty}$ on comparing coefficients of
equal charge. 

Both $\g FFF$ and $F\F\F$ terms in $\cl$ are of higher order in~$\a'$
and have the homogeneity of the generalised Leznov functional $\cl^{-4}$, 
rather than the generalised Plebanski functional $\cl^{-8}$ 
entering the hyperspace Lagrangean~\gl{hyperaction}. Although the
contribution of the $F\F\F$ term to the generalised Leznov equation~\gl{L}
is basically the `curving' of the flat hyperspace Leznov equation, both terms
yield novel contributions to the generalised Plebanski equation~\gl{mP}.
These contributions are proportional to two topological densities: 
the square of the self-dual Weyl 
curvature $\ C^{\da\db\dg\dd}C_{\da\db\dg\dd}\ $ and the trace of the self-dual
field-strength squared $\ Tr\ F^{\da\db}F_{\da\db}\ $. 
The appearance of the latter term was actually foreseen by early considerations
of Marcus \cite{marcus}. 

The `stringy' $\a'$-dependent terms do not appear to afford a fully
hyperspace-covariant formulation, although these terms are of manifestly 
geometric character, being proportional to the second Chern class of the 
hyperspace structure bundle and that of the Yang-Mills bundle respectively. 
These terms have the structure of torsion contributions, 
with the $\pa^{+\dn}$-derivative of \gl{mP} taking the form
\be
\cd^{-\da} E^{--\dn}_\da\ =\ \a'\  T^{---\dn} \quad .
\ee

The full coupled system \gl{L} and \gl{mP} 
shares with each of the uncoupled equations a conserved-current form,
\be
\pa^{+\da} \cj^{(-3)}_\da\ =\ 0 \qquad,\qquad
\pa^{+\da} J^{(-5)}_\da\ =\ 0 \quad,
\ee
with the currents being expressible in terms of higher prepotentials 
$\cy^{(-4)}$ and $Y^{(-6)}$ thus:
\bea
\cj^{(-3)}_\da &=\ \pa^+_\da \cy^{(-4)} 
&=\ \pa^-_\da  \F^{--}\ 
- \fr12\, [ \pa^+_\da  \F^{--}\  ,\  \F^{--} ]
- \pa^+_\da \pa^{+\db} F^{----}\   \pa^+_\db \F^{--}\ 
\nonumber\\[6pt]
J^{(-5)}_\da &=\ \pa^+_\da Y^{(-6)} 
&=\ \pa^-_\da  F^{----}\   
- \fr12\, \pa^+_\da \pa^{+\db} F^{----}\   \pa^+_\db F^{----}
\nonumber\\[4pt]
&&\quad +\ \fr{\a'}{2}\ (\eta^-)^4\  
        \Tr\ \pa^+_{\da}\pa^{+\db}\F^{--}\ \pa^+_{\db}\F^{--} 
\nonumber\\[4pt] 
&&\quad -\ \fr{\g\a'}{2}\  (\eta^+)^{-4}\Bigl( 
             \pa^+_{\da}  \pa^{+\db}\pa^{+\dg}\pa^{+\dd}F^{----}\ 
             \pa^+_{\db}\pa^+_{\dg}\pa^+_{\dd}F^{----}\Bigr) \quad.
\eea
The action of $\pa^{-\da}$ on these equations yields wave equations for the
higher prepotentials  $\cy^{(-4)}$ and $Y^{(-6)}$. These have
conserved-current form as well, yielding, in turn, higher prepotentials in 
the fashion of the uncoupled systems \cite{leznov,bp}. The towers of
higher prepotentials also encode the dynamics of the higher spin fields.
However, here we shall not pursue the relationship between the description
they provide and that offered by the coefficients in the $\eta$-expansions
of the hyperspace fields $F^{----}$ and $\F^{--}$ of present interest.


\section{Truncated actions}

Just as for the `flat' pure-Yang-Mills picture album discussed in \cite{dl1},
there exist various consistent truncations of $\cl_\infty$ to systems of finite
numbers of fields. 

The `maximal' consistent truncation has the $9$ $f$-type fields with $|j|\le 2$
coupled to the $5$ $\f$-type fields with $|j|\le 1$. This collection of fields
is obtained by
Taylor-expanding $F^{----}$ in powers of $1/\eta^+$ and $\F^{--}$ in powers of 
$\eta^-$, and truncating at orders $(\eta^+)^{-8}$ and $(\eta^-)^4$ 
respectively. Inserting in \gl{hyperaction} yields a homogeneous Lagrangean for
a multiplet of 9 Plebanski fields coupled to 5 Lie-algebra-valued Leznov 
fields, with helicities ranging in half-integer steps from ${+}2$ to ${-}2$  
and ${+}1$ to ${-}1$ respectively.
Setting $\a'{=}1$ for simplicity, we obtain
\bea
\cl_{9+5} &=&  -\ f^{++++} \square f^{----}\ -\ f^{+++} \square f^{---}\ 
       -\ f^{++} \square f^{--}\ -\  f^{+} \square f^{-}\ -\ \fr12 f \square f\ 
\nonumber\\[6pt] &&
 +\ \fr12\ f^{++++} \pa^{+\da}\pa^{+\db} f^{----} \pa^+_\da \pa^+_\db f^{----}\
   +\       f^{+++} \pa^{+\da} \pa^{+\db} f^{---} \pa^+_\da \pa^+_\db f^{----}
\nonumber\\[6pt] &&   
   +\       f^{++} \pa^{+\da} \pa^{+\db} f^{--} \pa^+_\da \pa^+_\db f^{----}\
   +\ \fr12\  f^{++} \pa^{+\da} \pa^{+\db} f^{---} \pa^+_\da \pa^+_\db f^{---}
\nonumber\\[6pt] &&  
   +\       f^{+} \pa^{+\da} \pa^{+\db} f^{-} \pa^+_\da \pa^+_\db f^{----}\
   +\       f^{+} \pa^{+\da} \pa^{+\db} f^{--} \pa^+_\da \pa^+_\db f^{---}
\nonumber\\[6pt] &&
   +\ \fr12\ f\ \pa^{+\da} \pa^{+\db} f\ \pa^+_\da \pa^+_\db f^{----}\
   +\ \fr{\g}{2}\  f \pa^{+\da}\pa^{+\db}\pa^{+\dg}\pa^{+\dd} f^{----}
                 \pa^+_{\da}\pa^+_{\db}\pa^+_{\dg}\pa^+_{\dd} f^{----}
\nonumber\\[6pt] &&  
   +\ \fr12\ f\ \pa^{+\da} \pa^{+\db} f^{--} \pa^+_\da \pa^+_\db f^{--}\
   +\ {\g}\  f^- \pa^{+\da}\pa^{+\db}\pa^{+\dg}\pa^{+\dd} f^{---}
                 \pa^+_{\da}\pa^+_{\db}\pa^+_{\dg}\pa^+_{\dd} f^{----}
\nonumber\\[6pt] &&
   +\       f\ \pa^{+\da} \pa^{+\db} f^{-}\ \pa^+_\da \pa^+_\db f^{---}\
   +\  \fr{\g}{2}\  f^{--} \pa^{+\da}\pa^{+\db}\pa^{+\dg}\pa^{+\dd} f^{--}
                        \pa^+_{\da}\pa^+_{\db}\pa^+_{\dg}\pa^+_{\dd} f^{----}
\nonumber\\[6pt] &&
  +\ \fr12\ f^{-}\ \pa^{+\da} \pa^{+\db} f^{-}\ \pa^+_\da \pa^+_\db f^{--}\   
       +\  \fr{\g}{2}\  f^{--} \pa^{+\da}\pa^{+\db}\pa^{+\dg}\pa^{+\dd} f^{---}
                        \pa^+_{\da}\pa^+_{\db}\pa^+_{\dg}\pa^+_{\dd} f^{---}
\nonumber\\[6pt] &&
+\  \Tr\ \biggl\{-\ \f^{++} \square \f^{--}\ -\ \f^{+} \square \f^{-}\
               -\fr12\ \f \square \f\ 
\nonumber\\[6pt] &&\qquad\quad
 + \left( \f^{++}\e^{\db\da}\ -\ \fr12\ \pa^{+\da} \pa^{+\db} f \right) 
                                           \pa^+_\da \f^{--} \pa^+_\db \f^{--}\
\nonumber\\[6pt] &&\qquad\quad
 + \left( 2\f^{+}\e^{\db\da}\ -\ \pa^{+\da} \pa^{+\db} f^{-} \right) 
                                           \pa^+_\da \f^{-} \pa^+_\db \f^{--}\
\nonumber\\[6pt] &&\qquad\quad
 + \left( \f\ \e^{\db\da}\ -\ \fr12\ \pa^{+\da} \pa^{+\db} f^{--} \right) 
                                           \pa^+_\da \f^{-} \pa^+_\db \f^{-}\
\nonumber\\[6pt] &&\qquad\quad
 + \left( \f\ \e^{\db\da}\ -\ \pa^{+\da} \pa^{+\db} f^{--} \right) 
                                           \pa^+_\da \f\ \pa^+_\db \f^{--}\
\nonumber\\[6pt] &&\qquad\quad
  -\ \pa^{+\da} \pa^{+\db} f^{---}  \left(
 \pa^+_\da \f^{+} \pa^+_\db \f^{--}\ +\ \pa^+_\da \f\ \pa^+_\db \f^{-} \right)
\nonumber\\[6pt] &&\qquad\quad
-\ \pa^{+\da} \pa^{+\db} f^{----}  \left(
  \pa^+_\da \f^{++} \pa^+_\db \f^{--}\ +\ \pa^+_\da \f^{+} \pa^+_\db \f^{-}\  
  +\ \fr12\  \pa^+_\da \f\ \pa^+_\db \f  \right)
\biggr\}\ .
\la{max}\eea
This truncated Lagrangean is remarkable in that it describes a 
{\it one-loop exact} theory; it is not hard to see that its Feynman 
rules do not support higher-loop diagrams. Further, this is the largest 
consistent subtheory of  $\cl_\infty$  with this property and with finitely 
many fields. Any attempt to include further fields necessarily requires 
the inclusion of the infinite set in order to obtain a consistent Lagrangean, 
and $\cl_\infty$ does not forbid multi-loop diagrams. 
The `flat limit', with the nine Plebanski-type fields $f^{\dots}$ set to zero, 
yields the five-field one-loop exact theory presented previously~\cite{dl1}.
The equations of motion for the Plebanski tower are
\bea
&\square f^{----} 
        &=\ \fr12\ \pa^{+\da} \pa^{+\db} f^{----} \pa^+_\da \pa^+_\db f^{----}
\nonumber\\[12pt] 
&\square f^{---}\ &=\ \pa^{+\da}\pa^{+\db} f^{---} \pa^+_\da \pa^+_\db f^{----}
\nonumber\\[12pt] 
&\square f^{--}\quad &=\ \pa^{+\da}\pa^{+\db}f^{--}\pa^+_\da\pa^+_\db f^{----}\
        +\ \fr12\ \pa^{+\da} \pa^{+\db} f^{---} \pa^+_\da \pa^+_\db f^{---}
\nonumber\\[12pt] 
&\square f^{-}\quad\ &=\ \pa^{+\da}\pa^{+\db}f^{-}\pa^+_\da\pa^+_\db f^{----}\
        +\ \pa^{+\da} \pa^{+\db} f^{--} \pa^+_\da \pa^+_\db f^{---}
\nonumber\\[12pt] 
&\square f\quad\quad\ &=\ \pa^{+\da}\pa^{+\db} f\ \pa^+_\da\pa^+_\db f^{----}\
        +\ \fr12\ \pa^{+\da} \pa^{+\db} f^{--} \pa^+_\da \pa^+_\db f^{--}
 \nonumber   \\[4pt]&&\quad        
        +\ \pa^{+\da} \pa^{+\db} f^{-} \pa^+_\da \pa^+_\db f^{---}
        +\ \fr{\g}{2}\ \pa^{+\da}\pa^{+\db}\pa^{+\dg}\pa^{+\dd} f^{----}
                       \pa^+_{\da}\pa^+_{\db}\pa^+_{\dg}\pa^+_{\dd} f^{----}
 \nonumber   \\[4pt]&&\quad    
   -\ \Tr\ \fr12\ \pa^{+\da} \pa^{+\db} \f^{--}\ \pa^+_\da \pa^+_\db \f^{--}
\nonumber\\[12pt] 
&\square f^{+} \quad\ 
             &=\ \pa^{+\da} \pa^{+\db} f^{+}\ \pa^+_\da \pa^+_\db f^{----}\
        +\ \pa^{+\da} \pa^{+\db} f^{-} \pa^+_\da \pa^+_\db f^{--}
 \nonumber   \\[4pt]&&\quad        
        +\ \pa^{+\da} \pa^{+\db} f\ \pa^+_\da \pa^+_\db f^{---}
        +\ {\g}\ \pa^{+\da}\pa^{+\db}\pa^{+\dg}\pa^{+\dd} f^{---}
                      \pa^+_{\da}\pa^+_{\db}\pa^+_{\dg}\pa^+_{\dd} f^{----} 
 \nonumber   \\[4pt]&&\quad        
        -\ \Tr\ \pa^{+\da} \pa^{+\db} \f^{-}\ \pa^+_\da \pa^+_\db \f^{--}
\nonumber  \\[12pt] 
&\square f^{++}\quad 
             &=\ \pa^{+\da} \pa^{+\db} f^{++}\ \pa^+_\da \pa^+_\db f^{----}\
        +\  \pa^{+\da} \pa^{+\db} f^{+}\ \pa^+_\da \pa^+_\db f^{---}            
 \nonumber   \\[4pt]&&\quad                  
        +\  \pa^{+\da} \pa^{+\db} f\  \pa^+_\da \pa^+_\db f^{--}
        +\ {\g}\ \pa^{+\da}\pa^{+\db}\pa^{+\dg}\pa^{+\dd} f^{--}
                      \pa^+_{\da}\pa^+_{\db}\pa^+_{\dg}\pa^+_{\dd} f^{----} 
 \nonumber   \\[4pt]&&\quad    
        +\  \fr12\  \pa^{+\da} \pa^{+\db} f^{-}\ \pa^+_\da \pa^+_\db f^{-} 
        +\  \fr{\g}{2}\   \pa^{+\da}\pa^{+\db}\pa^{+\dg}\pa^{+\dd} f^{---}
                      \pa^+_{\da}\pa^+_{\db}\pa^+_{\dg}\pa^+_{\dd} f^{---}\
 \nonumber   \\[4pt]&&\quad    
 -\ \Tr\ \left( \fr12\ \pa^{+\da} \pa^{+\db} \f^{-}\ \pa^+_\da \pa^+_\db \f^{-}
        +\ \pa^{+\da} \pa^{+\db} \f\ \pa^+_\da \pa^+_\db \f^{--} \right)
\nonumber  \\[12pt] 
&\square f^{+++}\ 
             &=\ \pa^{+\da} \pa^{+\db} f^{+++}\  \pa^+_\da \pa^+_\db f^{----}\
        +\  \pa^{+\da} \pa^{+\db} f^{++}\ \pa^+_\da \pa^+_\db f^{---}        
 \nonumber   \\[4pt]&&\quad                
        +\  \pa^{+\da} \pa^{+\db} f^{+}\  \pa^+_\da \pa^+_\db f^{--}
        +\  {\g}\   \pa^{+\da}\pa^{+\db}\pa^{+\dg}\pa^{+\dd} f^{-}
                      \pa^+_{\da}\pa^+_{\db}\pa^+_{\dg}\pa^+_{\dd} f^{----}
 \nonumber   \\[4pt]&&\quad    
        +\  \pa^{+\da} \pa^{+\db} f\ \pa^+_\da \pa^+_\db f^{-}
        +\  {\g}\   \pa^{+\da}\pa^{+\db}\pa^{+\dg}\pa^{+\dd} f^{--}
                      \pa^+_{\da}\pa^+_{\db}\pa^+_{\dg}\pa^+_{\dd} f^{---}\
 \nonumber   \\[4pt]&&\quad   
 -\ \Tr\ \left( \pa^{+\da} \pa^{+\db} \f^{+}\ \pa^+_\da \pa^+_\db \f^{--}
        +\ \pa^{+\da} \pa^{+\db} \f\ \pa^+_\da \pa^+_\db \f^{-} \right)
\nonumber  \\[12pt]  
&\square f^{++++} 
             &=\ \pa^{+\da} \pa^{+\db} f^{++++}\  \pa^+_\da \pa^+_\db f^{----}\
        +\  \pa^{+\da} \pa^{+\db} f^{+++}\ \pa^+_\da \pa^+_\db f^{---}        
 \nonumber   \\[4pt]&&\quad                
        +\  \pa^{+\da} \pa^{+\db} f^{++}\  \pa^+_\da \pa^+_\db f^{--}
        +\  \pa^{+\da} \pa^{+\db} f^{+}\ \pa^+_\da \pa^+_\db f^{-}
 \nonumber   \\[4pt]&&\quad  
        +\ \fr12\ \pa^{+\da} \pa^{+\db} f\ \pa^+_\da \pa^+_\db f\ 
        +\  \fr{\g}{2}\   \pa^{+\da}\pa^{+\db}\pa^{+\dg}\pa^{+\dd} f^{--}
                      \pa^+_{\da}\pa^+_{\db}\pa^+_{\dg}\pa^+_{\dd} f^{--}
 \nonumber   \\[4pt]&&\quad  
        +\ {\g}\ \pa^{+\da}\pa^{+\db}\pa^{+\dg}\pa^{+\dd} f^{-}
                      \pa^+_{\da}\pa^+_{\db}\pa^+_{\dg}\pa^+_{\dd} f^{---}  
 \nonumber   \\[4pt]&&\quad   
        +\ {\g}\  \pa^{+\da}\pa^{+\db}\pa^{+\dg}\pa^{+\dd} f
                      \pa^+_{\da}\pa^+_{\db}\pa^+_{\dg}\pa^+_{\dd} f^{----}
        -\ \Tr \left(\pa^{+\da} \pa^{+\db} \f^{++}\ \pa^+_\da \pa^+_\db \f^{--}
\right.\nonumber \\[4pt] && \left.\quad
        +\ \pa^{+\da} \pa^{+\db} \f^{+}\ \pa^+_\da \pa^+_\db \f^{-}
        +\ \fr12\ \pa^{+\da} \pa^{+\db} \f\ \pa^+_\da \pa^+_\db \f\ \right)
              \quad.
\la{9}\eea

The five fields of the Leznov tower satisfy curved space versions of the 
five flat space equations given in \cite{dl1}, 
\bea
&\square \f^{--} &=\
           \fr12 [ \pa^{+\da} \f^{--} , \pa^+_\da \f^{--} ]\ 
        +\ \pa^{+\da} \pa^{+\db} f^{----} \pa^+_\da \pa^+_\db \f^{--}
\nonumber\\[12pt]
&\square \f^{-}\ &=\
            [ \pa^{+\da} \f^{-} , \pa^+_\da \f^{--} ] 
        +\ \pa^{+\da} \pa^{+\db} f^{---} \pa^+_\da \pa^+_\db \f^{--}\  
\nonumber\\[4pt] && \quad    
       +\ \pa^{+\da} \pa^{+\db} f^{----} \pa^+_\da \pa^+_\db \f^{-}
\nonumber\\[12pt]
&\square \f\quad &=\
            [ \pa^{+\da} \f , \pa^+_\da \f^{--} ]\ +\
            \fr12 [ \pa^{+\da} \f^{-} , \pa^+_\da \f^{-} ] 
\nonumber\\[4pt] &&\quad    
       +\ \pa^{+\da} \pa^{+\db} f^{--} \pa^+_\da \pa^+_\db \f^{--}\           
       +\ \pa^{+\da} \pa^{+\db} f^{---} \pa^+_\da \pa^+_\db \f^{-}    
\nonumber\\[4pt] &&\quad      
       +\ \pa^{+\da} \pa^{+\db} f^{----} \pa^+_\da \pa^+_\db \f 
\nonumber\\[12pt]       
&\square \f^{+}\ &=\
            [ \pa^{+\da} \f^{+} , \pa^+_\da \f^{--} ]\ +\
            [ \pa^{+\da} \f , \pa^+_\da \f^{-} ] 
\nonumber\\[4pt]&&\quad   
       +\ \pa^{+\da} \pa^{+\db} f^{-} \pa^+_\da \pa^+_\db \f^{--}\           
       +\ \pa^{+\da} \pa^{+\db} f^{--} \pa^+_\da \pa^+_\db \f^{-}    
\nonumber\\[4pt] &&\quad      
       +\ \pa^{+\da} \pa^{+\db} f^{---} \pa^+_\da \pa^+_\db \f\ 
       +\ \pa^{+\da} \pa^{+\db} f^{----} \pa^+_\da \pa^+_\db \f^{+}
\nonumber\\[12pt]       
&\square \f^{++} &=\
            [ \pa^{+\da} \f^{++} , \pa^+_\da \f^{--} ]\ +\
            [ \pa^{+\da} \f^{+} , \pa^+_\da \f^{-} ]\ +\
            \fr12 [ \pa^{+\da} \f , \pa^+_\da \f ] 
\nonumber\\[4pt]&&\quad  
       +\ \pa^{+\da} \pa^{+\db} f\ \pa^+_\da \pa^+_\db \f^{--}\           
       +\ \pa^{+\da} \pa^{+\db} f^{-} \pa^+_\da \pa^+_\db \f^{-}    
\nonumber\\[4pt] &&\quad     
       +\ \pa^{+\da} \pa^{+\db} f^{--} \pa^+_\da \pa^+_\db \f\ 
       +\ \pa^{+\da} \pa^{+\db} f^{---} \pa^+_\da \pa^+_\db \f^{+}   
\nonumber\\[4pt] &&\quad     
       +\ \pa^{+\da} \pa^{+\db} f^{----} \pa^+_\da \pa^+_\db \f^{++}            
            \quad.
\la{5}\eea
Picture-raising induces a derivation $Q^+$ on the set of target space fields,
\be\arr
&&Q^+:\ f^{----} \mapsto f^{---} \mapsto 2\ f^{--}
\mapsto 3!\ f^{-} \mapsto\ 4!\  f
\mapsto 5!\ f^+ 
\mapsto\dots 
\mapsto 8!\ f^{++++}\\[8pt]
&&Q^+:\ \f^{--}\ \mapsto\ \f^-\ \mapsto\ 2\ \f\
\mapsto\ 3!\ \f^+\ \mapsto\ 4!\ \f^{++} \quad.
\ea\ee
The five-equation system \gl{5} follows from the $\f^{--}$ equation on
successive application of $Q^+$. This property, displayed for the corresponding
flat space equations in \cite{dl1}, therefore survives the coupling to the
five $f$-type fields occurring in  \gl{5}. For the nine-equation tower \gl{9},
successive application of $Q^+$ on the `top' ($f^{----}$) equation yields
all the `non-stringy' terms, namely those {\it not} depending on the
suppressed $\a'$. On the other hand, these `stringy' terms follow on successive 
application of $Q^+$ to the two topological densities inserted in the neutral 
$f$ equation. However, the relative normalisations of these two sets of terms
are not suited to the consideration of the $f$ equation as the
`top' equation for the positively charged equations.

The  fourth-order ($\g$-dependent) `stringy' terms of the Lagrangean 
$\,\cl_{9+5}\,$ are seen to affect only the equations for the gravitational 
`multiplier' fields~$f_{j\le0}\,$ and do not enter the Plebanski equations for 
the positive-helicity (negatively-charged) fields. 
In the high string tension limit, $\a'\to 0$, these terms in any case disappear
and we recover equations which arise on expanding \gl{P}.
The above $9{+}5$ field system, apart from the $\a'$-dependent deformation,
is indeed somewhat similar to self-dual \N8  {\it super\/}gravity 
\cite{siegel_sdsg} plus $N{=}4$ self-dual {\it super\/} Yang-Mills 
\cite{siegel_sdym,CS}, with adjustments made for the 
difference in statistics of the spinorial fields. We note however, that 
whereas the five fields of the $\F^{--}$ multiplet are in one-to-one
correspondence with the components of the \N4 SDYM multiplet \cite{dl1},
the \N8 supermultiplet of \cite{siegel_sdsg} has eleven component fields.

Starting from the $9{+}5$ field truncation, smaller consistent Lagrangean
theories may be constructed by ignoring any selection of pairs of fields from
$\{ (f_j , f_{-j}),(\f_j , \f_{-j})\}$.  Any such 
truncation of the `maximal model' may easily be seen to be one-loop finite.  
We note that the `minimal model', containing only the standard Plebanski and 
Leznov fields, $f^{----}$ and $\f^{--}$, together with their respective 
multipliers, $f^{++++}$ and $\f^{++}$, does not even contain a `$\g$-term'.


\section{Conclusions}

We have seen that the classical curved-space self-duality equations 
in $(2,2)$ hyperspace describe the interaction of 
open and closed $N{=}2$ strings, at least on topologies with $\c{>}0\,$,
up to stringy torsion-like modifications which vanish in the 
high tension limit. Since massive $N{=}2$ string excitations do not exist,
these $\a'$-corrections are actually unexpected. They owe their appearance
to the picture degeneracy of the massless level, which also forms
the basis of the hyperspace extension of self-duality.
The second order $\a'$-terms, moreover, are seen to be indispensable
for the formulation of a unified action principle
for the coupled self-dual Einstein-Yang-Mills system.

In the hyperspace formulation of our coupled system \gl{hyperaction}, the
stringy modifications depend explicitly on the spinorial hyperspace coordinates
($\eta^\pm$) and do not seem to afford a fully hyperspace-covariant
reformulation. This difficulty is actually related to the `wrong' statistics of
the spinorial coordinate.  In {\it super\/}space, the difference in
dimension of the two integration measures, $d^4\q$ for the \ym\ terms and
$d^8\q$ for the gravitational ones, makes it possible to construct a covariant
combined action.

The existence of the higher conserved currents and potentials (section 4)
seems to indicate that the $\a'$-deformation introduced here does not
affect the integrability of the coupled model. The full system described 
by \gl{Linfty} therefore deserves further study in this light. 
Relaxing the string-enforced requirement of a fixed complex structure,
we may recover full Lorentz invariance in harmonic space, with the
$\{u^\pm_\a\}$ of section 2 treated as genuine coordinates. It remains to be
seen whether such a reformulation provides an  $\a'$-deformation of the 
Penrose twistor transform.


\noindent
{\bf Acknowledgment} 

\noindent
We thank J\"urgen Schulze for discussions concerning the relation
of string theory to field theory amplitudes.


\end{document}